\documentclass[12pt,preprint]{aastex}
 \usepackage{epsf}
 \usepackage{amsmath}
 \usepackage[mathscr]{eucal}
 \usepackage{mathrsfs} 

\let\DS=\displaystyle

\newcommand{\EXP}[1]{{\rm e}^{#1}}
\newcommand{\mathe}{\mathrm{e}}

\newcommand{\D}{{\rm d}}
\newcommand{\para}[1]{{\left(#1\right)}}

\newcommand{\vC}{$\check{\rm C}$}

\newcommand{\mscr}[1]{\bgroup\mathscr #1\egroup}
\slugcomment{Accepted for publication in the Astrophysical Journal}
\shorttitle{\uppercase{Cooling `pile-up' EDF gamma-ray emission}}
\shortauthors{\uppercase{Saug\'e\ \&\ Henri}}
\begin{document}
\title{
   T\lowercase{e}V blazar gamma-ray emission produced by \\   
   a cooling pile-up particle energy distribution function    
}
\author{%
	Ludovic Saug\'e
             and
	Gilles Henri
}
\affil{Laboratoire d'Astrophysique de Grenoble, Universit\'e
   Joseph-Fourier, BP~53, F-38041 Grenoble, France
}
\email{Ludovic.Sauge@obs.ujf-grenoble.fr, Gilles.Henri@obs.ujf-grenoble.fr}
\begin{abstract}
We propose a time-dependent one-zone model based on a quasi-Maxwellian
`pile-up' distribution in order to explain the time-averaged high energy
emission of TeV blazars. The instantaneous spectra are the result of the
synchrotron and synchrotron self-Compton emission (SSC) of ultra-relativistic
leptons. The particle energy distribution function (EDF) is computed in a
self-consistent way, taking into account an injection term of fresh particles,
a possible pair creation term, and the particles radiative cooling. The source
term is not a usual power-law but rather a `pile-up' distribution, which can
result  from the combination of a stochastic heating via second order Fermi
process and radiative cooling. To validate this approach, we have performed
time-averaged fits of the well-known TeV emitter Mrk\ 501 during the 1997
flaring activity period taking into account the attenuation of the high energy
component by cosmic diffuse infrared background (DIrB) and intrinsic absorption
via the pair creation process. The model can reproduce very satisfactorily the
observed spectral energy distribution (SED). A high Lorentz factor is required
to avoid strong pair production; in the case of smaller Lorentz factor, an
intense flare in the GeV range is predicted due to the sudden increase of soft
photons density below the Klein-Nishina threshold. The possible relevance of
such a scenario is discussed.
\end{abstract}
\section{Introduction}
It is now widely admitted that radio-loud active galactic nuclei (AGN)
harbor magnetized accretion-ejection structures involving a supermassive
black-hole as a central engine. The EGRET experiment aboard the
Compton-GRO satellite discovered more than 80 gamma-ray emitting AGNs,
all of them belonging of the blazar class (non-thermal continuum
spectrum, optical polarization, flat radio spectrum and strong
variability in all frequency bands). Some of these objects have been
also firmly detected by Atmospheric {\vC}erenkov Telescope (ACT) with an
emission above 1 TeV. The two prototypes of TeV blazars are Mrk\ 421
\citep{punch92} and Mrk\ 501 \citep{quinn96}, two objects relatively
close to us and roughly at the same distance, respectively
$z_s\sim0.031$ and $z_s\sim0.034$. Thanks to the development of the
ground-based gamma-ray astronomy the sample of the TeV emitters is
increasing. During the last decade, several ACT teams have reported the
detection or the confirmation of new sources : 1ES\,1426+428
\citep{horan02,djannati02,aharonian02}, 1ES\,1959+650
\citep{nishiyama00,aharonian03,holder03}, 1ES\,2344+514
\citep{catanese98} and PKS\,2155-304 \citep{chadwick99}.
A characteristic feature of blazars is the strong non-thermal emission
from the radio to the gamma-ray range attributed to a relativistic jet
supposed to be closely aligned with the observer light-of-sight. Their
spectral energy distribution is quite typical and consists in two broad
bumps. In the context of the Synchrotron Self-Compton (SSC) models, the
low energy component, peaking in the X-ray domain for TeV blazars, is
commonly attributed to the synchrotron emission of ultra-relativistic
particles plunged into magnetic field.  The second one is thought to be
the result of up-scattering of the synchrotron photon field by the same
population of ultra-relativistic particles via the inverse Compton (IC)
mechanism \citep{jones74,konigl81,ghisellini85}.  This model gives a
good framework to explain the correlated variability for the the high
and the low energy components.
Even if the spectral properties of these objects seem to be understood,
the different models do not discuss the origin and the physical
mechanism of particle acceleration. To reproduce the curved shape of the
synchrotron and IC spectra on a wide energy domain, several authors have
chosen a particle EDF parameterized by a simple or a broken power-law on
a prescribed energy range $[\gamma_{\rm min},\gamma_{\rm max}]$. This
choice is purely phenomenological and have no theoretical justification,
even if in some special cases of shock acceleration (first order Fermi
process) power-law EDF are expected \citep{jones94}. For example, to
reproduce the X-ray synchrotron bump, several authors use a simple
power-law $n(\gamma)\propto\gamma^{-s}$, $\gamma\in[\gamma_{\rm
min},\gamma_{\rm max}]$ but the dynamical range \textit{i.e.} the ratio
of $\gamma_{\rm max}/\gamma_{\rm min}$ is less than 10 \citep{pian98}.
In this case, it seems to be more appropriate to consider a quasi
mono-energetic distribution.
In this work, we propose another primary type of EDF for emitting
particles in order to reproduce the peculiar spectral energy
distribution of TeV blazars. We assume that the acceleration mechanism
combined to radiative losses or/and an escape process produces a
quasi-Maxwellian or `pile-up' distribution, which is injected in a
spherical region where it cools freely.  The effect of cooling is to
produce naturally a $E^{-2}$ power-law in some limited range of energy.
We also take into account the time dependence of the EDF to compare with
the observations, considering that the observed spectra are always
time-averaged spectra of intrinsically highly variable objects.  In
section 2, we present our kinetic scenario to obtain the energy spectrum
of the particles in a self-consistent way, and we shortly describe the
emission processes used to reproduce the blazar spectra. Finally we
illustrate our approach in section 3, giving some results of SED fitting
before concluding.
\section{The model}
\subsection{Stochastic particles acceleration}
In the following, we will consider only a homogeneous one-zone model
where all physical quantities are assumed to be averaged over the volume
of the emission region. All spatial dependences are dropped from the
equations. The particles distribution function $f(\pmb{p};t)$ is assumed
to be isotropic in some frame, called the `blob frame', moving
relativistically with a bulk Lorentz factor $\Gamma_B$. In this frame,
it depends only on the modulus of the momentum $p=|\pmb{p}|$ and the
time $t$. For relativistic particles the energy is given by $E=\gamma
m_{\rm e}c^2\sim pc$ and the differential number density of pairs
$n_{\pm}(\gamma;t)$ of reduced energy $\gamma$ is related to EDF
$f(\gamma)$ by the usual relation (time is implicit) ${\rm
d}n_{\pm}=n_{\pm}(\gamma)\,{\rm d}\gamma= 4\pi p^2f(p)\,{\rm d}p
\sim4\pi (m_{\rm e}c^2)^3 \gamma^2 f(\gamma)\,{\rm d}\gamma $.

We assume that the particles are accelerated stochastically by energy
exchanges with resonant plasma waves in a weak turbulent medium.  In our
model, the acceleration zone must be localized : it could be the basis
of a jet, or localized reconnection sites, or the interface between a
relativistic beam and a confining jet  as proposed for example by
\citet{hp91} in the framework of the `two-flow model'
\citep{pel85,pel92}.  This insures that the particles will spend only a
tiny fraction of time in the acceleration zone, before being injected in
a larger region where they cool freely.  According to quasi-linear
theory, the acceleration process can be described by a diffusion
equation in the momentum space leading to a Fokker-Planck equation. This
equation gives the time-dependent evolution of any initial particle
density submitted to deterministic continuous energy changes or
diffusive Markovian processes. We suppose that the characteristic
acceleration time-scale is short compared with the other time-scales in
the problem and we will focus our attention onto the stationary solution
$f(\gamma)$ of the Fokker-Planck equation.
The diffusion coefficient $\mathscr D_{\gamma\gamma}(\gamma)$ in phase
space can be chosen as a power-law in terms of the Lorentz factor
$\gamma$ \citep{lacombe77,hp91,dermer96}
  \begin{equation*}
  \mathscr D_{\gamma\gamma}(\gamma)=\mathscr D_0 \gamma^r,
  \end{equation*}
where $r\in[1,2]$ is the index of wave turbulent spectrum , assumed to
be itself a power-law (\textit{e.g.}  $r=5/3$ for a Kolmogorov
turbulence, $r=3/2$ for Kraishman one). 
The steady-state differential energy spectrum resulting from a
competitive balance between usual radiative cooling processes and
stochastic acceleration is a relativistic Maxwellian function also
called `pile-up' distribution \citep{schlickeiser85,aharonian86,hp91},
  \begin{equation}\label{eq:pileup}
  n(\gamma)\propto
  	\gamma^2\,\exp \left[
  	- \para{\gamma \over \gamma_{\rm max}}^{3-r}
  	\right],
  \end{equation}
and where $\gamma_{\rm max}$ is simply the value of the individual
Lorentz factor of the particles for which the acceleration time is equal
to the cooling time. It corresponds to an energy distribution function
of particles homogeneous and isotropic in the momentum space with a
exponential cut-off at $\gamma_{\rm max}$. Note that in the case of a
power-law distribution function (with spectral index $s>2$), the
enthalpy of the plasma is dominated by lower bound of the particle
energy range $\gamma_{\rm min}$. For a `pile-up', particles are mostly
concentrated near $\gamma_{\rm max}$ and the dynamics of the plasma is
mainly controlled by the high energy particles.\\
The inclusion of an escape term will modify the above solution. The model presented here
will break down if the escape time is much smaller or much larger than the characteristic acceleration 
time at the critical Lorentz factor $\gamma_{\rm max}$.  In the first case, acceleration will be much slower than the escape
and no relativistic pile-up can be formed. In the second case, the relativistic particles will remain a long time before escaping (and cooling) and the emission of the acceleration zone will be important. In the following, we exclude these two cases and we assume that the escape time is comparable to other times at  $\gamma_{\rm max}$, neglecting the emission of the acceleration zone. A proper inclusion of the escape term would modify the solution of the type
given by equation (\ref{eq:pileup}), but the general shape would be the same ; a low energy part behaving like $\gamma^2$ when the acceleration/diffusion is very fast, followed by an energy cut-off. For sake of simplicity, we will thus use equation (\ref{eq:pileup}) and replace 
the $3-r$ exponent by 1. The shape of the SED high energy tail is only
weakly dependent on this approximation, and not strongly constrained by
the observations. \\
\subsection{The cooling zone}
In order to obtain the energy spectrum of emitting particles, we assume
that the particles are accelerated as previously described in some
localized region and are injected during some time in a spherical zone
where they cool freely. In this zone, we consider the standard kinetic
equation in the continuous loss approximation with no escape term. It
gives the evolution of the differential energy density of the particles
$n_{\pm}(\gamma;t)$ with a Lorentz factor between $\gamma$ and
$\gamma+{\rm d}\gamma$,
\begin{equation}\label{eq:kin}
\dfrac{\partial}{\partial
t}n_{\pm}(\gamma;t)+\dfrac{\partial}{\partial
\gamma}\dot\gamma(\gamma;t)n_{\pm}(\gamma;t)=Q(\gamma;t).
\end{equation}
Particles source term $Q(\gamma;t)$ will include in fact both the fresh
particles injection term $Q_{\rm inj}(\gamma;t)$  and the production
rate $Q_{\rm prod}(\gamma;t)$ due to the pair creation via photon-photon
annihilation, which will be developed in section \ref{sec:pprat}. We
take the following approximate form for the injection term :
\begin{equation}\label{eq:injectionterm}
Q_{\rm inj}(\gamma;t)=
\begin{cases}
   n_0\gamma^2\exp (-{\gamma/\gamma_{\max}}) &\text{if } 0\leqslant t 
\leqslant t_{\rm inj},\\
0 & \text{otherwise,}
\end{cases}
\end{equation}
The factor $\dot\gamma(\gamma;t)$ in the energy advective part of
equation (\ref{eq:injectionterm}) is the continuous particle cooling
rate.  As mentioned above, charged particles can cool both via the
synchrotron process or via the IC scattering of the previous synchrotron
radiation field. We can thus write :
\begin{equation}\label{eq:coolingrate}
\dot\gamma(\gamma;t)=\dot\gamma_{\rm syn}(\gamma;t)+\dot\gamma_{\rm 
IC}(\gamma;t).
\end{equation}
\subsection{The radiative processes}\label{sec:radpro}
In the following we detail the equations used to compute the radiative
processes.  A `tilde' accent denotes a parameter expressed in the
observer frame, otherwise in the blob frame.
\subsubsection{The synchrotron emission}\label{sec:synchprocess}
Assuming an isotropic particle distribution, the synchrotron cooling
rate is  given by the well-known formula:
\begin{equation}\label{eq:syncoolingrate}
\dot\gamma_{\rm syn}(\gamma;t) = -k_{\rm syn} \gamma^2 \quad;\quad k_{\rm 
syn}=\dfrac{4}{3}\dfrac{\sigma_{\rm Th}}{m_ec}U_B,
\end{equation}
where $U_B=B^2/8\pi$ is the magnetic energy density.
The synchrotron emission coefficient $j_{\rm S}(\nu)$ is obtained by
performing the integration over the whole differential particle density
of the mean emission coefficient for a single lepton averaged over an
isotropic distribution of pitch angles $R_{\rm CS}(z)$
\citep{crusius86,ghisellini88}
\begin{equation}
   j_{\rm S}(\nu;t) = \dfrac{\sqrt3 e^3 B}{4\pi m_ec^2}
                    \int {\rm d}\gamma\,n_\pm(\gamma;t)\,R_{\rm CS}(z),
\end{equation}
\begin{equation}\label{eq:RCS}
   R_{\rm CS}(z) = 2z^2
   \left[
     K_{4/3}(z)K_{1/3}(z)-\dfrac {3z}5
     \left(
     K_{4/3}^2(z)-K_{1/3}^2(z)
     \right)
     \right],
\end{equation}
with $z={\nu}/{3\gamma^2\nu_B}$ and $\nu_B={eB}/{2\pi m_ec}$, $K_n$
being the McDonald function of order $n$. An accurate approximation of
the function $R_{\rm CS}(z)$ is given in appendix \ref{ap:synch}. 
\subsubsection {The Inverse Compton emission}
In the same way, the Inverse Compton scattering cooling rate reads
\begin{equation}\label{eq:ICcoolingrate}
   \dot\gamma_{\rm IC}(\gamma;t) =
   \int\D\epsilon_1\,\epsilon_1 \int\D \epsilon\, {\mathcal K}_{\rm 
Jones}(\epsilon_1,\epsilon,\gamma)\,
   n_{\rm syn}(\epsilon;t),
\end{equation}
where we consider the Compton kernel $\mathcal K_{\rm Jones}$ computed
by Jones \citep{jones68,bg70} for an isotropic source of soft photons,
considering the full Klein-Nishina cross section in the head-on
approximation. More precisely, we have
\begin{gather}\label{eq:jones}
    \mathcal K_{\rm Jones}(\epsilon_1,\epsilon,\gamma) =
    \dfrac 34 \dfrac{ c\sigma_{\rm Th}}{\epsilon\gamma^2} f(q,\Gamma_\epsilon)
    \Theta(q-1/4\gamma^2)\Theta(1-q),\\
    \Gamma_\epsilon=4\epsilon\gamma,
    \quad q=\dfrac{\epsilon_1}{4\epsilon\gamma(\gamma-\epsilon_1)},\\
    f(q,\Gamma_\epsilon) =
    \left[
      2q\ln q +(1+2q)(1-q)+\dfrac 12 \dfrac{(\Gamma_\epsilon 
q)^2}{1+\Gamma_\epsilon q}(1-q)
      \right],
  \end{gather}
where $\Theta(t)$ is the usual Heaviside one-step-function and
$\sigma_{\rm Th}$ the Thomson cross section.  We derive the IC emission
coefficient $j_{\rm C}(\nu)$ in a similar way. The Compton kernel
(equation \ref{eq:jones}) is this time integrated over the synchrotron
emission spectrum :
\begin{equation}\label{eq:jssc}
   j_{\rm C}(\nu_1;t) = \dfrac{h}{4\pi}\epsilon_1
   \iint {\mathcal K}_{\rm Jones}(\epsilon_1,\epsilon,\gamma)\,
   n_{\rm syn}(\epsilon;t) n_\pm(\gamma;t)\,
   {\rm d}\gamma{\rm d}\epsilon,
\end{equation}
where $n_{\rm syn}(\epsilon)$ is the differential synchrotron photon
density and $\epsilon_1=h\nu_1/m_ec^2$. These equations are integrated
numerically following the time evolution of the particle spectrum to
find the time-dependent emission spectrum of the plasma.
\subsubsection{The pair creation process}\label{sec:pppro}
\begin{figure}
\plotone{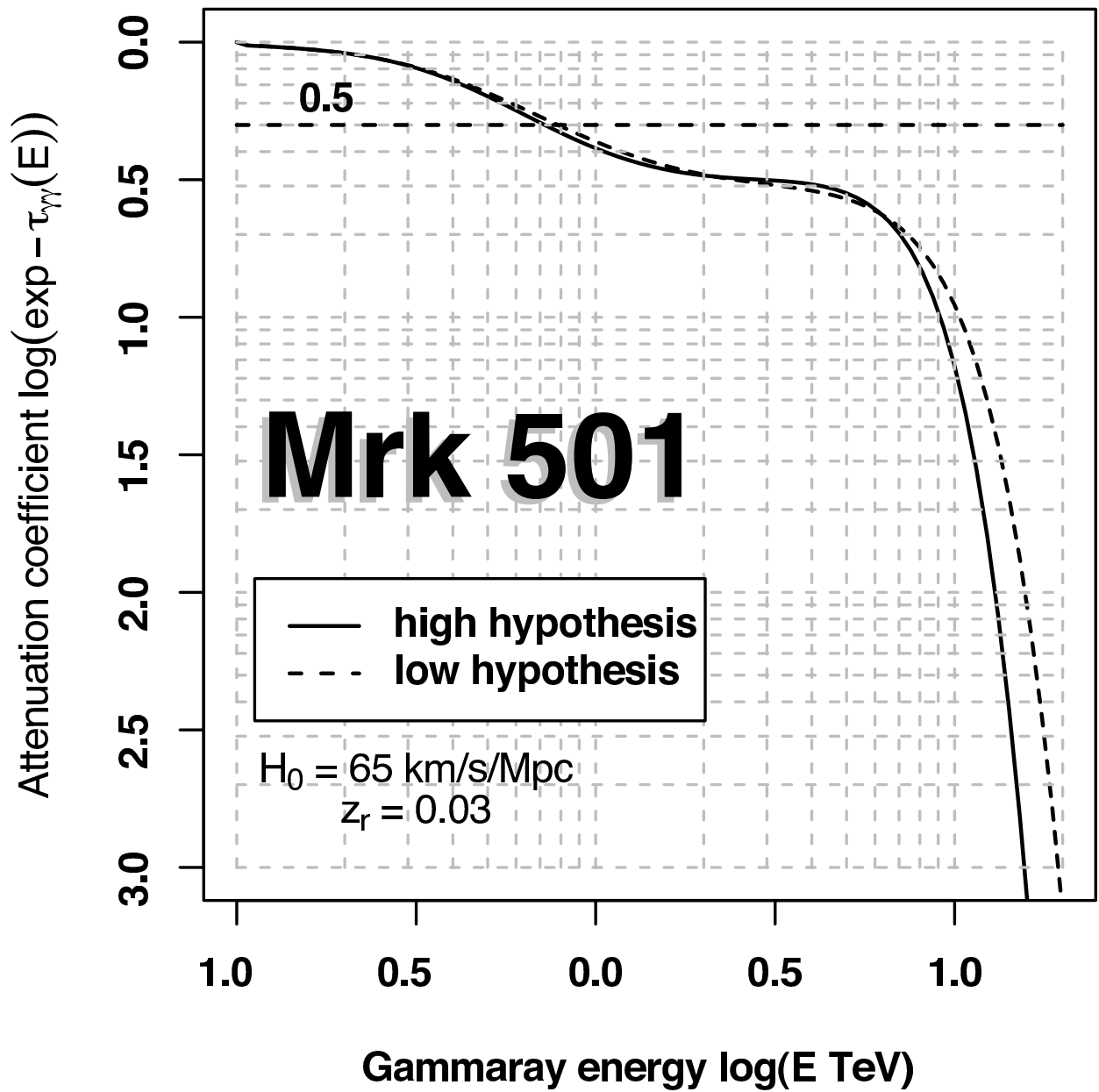}
\caption{%
         Synthetic absorption coefficients calculated for Mrk\ 501
         ($z_s\sim0.034$, $H_0=65\ \rm km\,s^{-1}\,Mpc^{-1}$). Solid
         curve (\emph{high hypothesis}) takes into account the 100 and
         60$\ \mu$m flux reported where dashed curve (\emph{low
         hypothesis}) does not.
	}\label{fig:taugg}
\end{figure}

Gamma-rays photons produced in the blob can be absorbed by the
photo-annihilation/pair creation process $\gamma+\gamma\to{\rm
e}^{+}+{\rm e}^{-}$ \citep{gould67a} for which the total cross section
is
\begin{equation}
   \sigma(x) =\frac{3\sigma_{\rm Th}}{16} (1-x^2)
   \left[
     (3-x^4)\ln\dfrac{1+x}{1-x}-2x(2-x^2)
     \right],
\end{equation}
The $\gamma$-$\gamma$ attenuation optical depth per unit length $\D \ell$
reads 
\begin{equation}\label{eq:taugg}
  \dfrac{\rm d}{\rm d\ell}\tau_{\gamma\gamma}(\epsilon_1)=
   \iiint \sigma(\beta)n_{\rm ph}(\epsilon_2,\Omega)(1-\mu){\rm 
d}\epsilon_2{\rm d}^2\Omega,
\end{equation}
where  $ \beta \equiv \beta(\epsilon_1,\epsilon_2,\mu) = \left(
1-2/\epsilon_1\epsilon_2(1-\mu) \right)^{1/2} $ is the velocity of the
pairs in the center-of-mass system, $\epsilon_1$ (resp. $\epsilon_2$)
the energy of the high (resp. low) energy photon and $\mu$ the cosine of the
collision angle \citep{coppi90}.
For gamma-rays in the TeV range the pair production cross section is
maximized when the soft photon energy is in the infrared range,
\begin{equation}\label{eq:interactrel}
\lambda
=\lambda_{ce}\dfrac{E_{\gamma}}{2 m_ec^2}
\sim 2.4\ \dfrac{E_{\gamma}}{1{\text{ TeV}}}\ \mu{\rm m},\quad \text{where }
\lambda_{ce}=\dfrac{h}{m_ec}.
\end{equation}
According to the previous relation, we can distinguish two different
sources of soft photons being able to absorb the high energy tail of
blazars.
Firstly, through \emph{intrinsic attenuation},  gamma-ray photons can
interact with photons of the synchrotron bump in the source. Assuming
again the synchrotron photon field to be isotropic in the blob frame,
the integration over solid angle in equation (\ref{eq:taugg}) can be 
analytically computed. More precisely, one gets :
\begin{equation}
   \dfrac{{\rm d}}{{\rm d}\ell}\tau_{\gamma\gamma}(\epsilon_1) = \frac 1c\int
   n_{\rm phX}(\epsilon_2) \mathscr R_{\rm 
pp}(\epsilon_1\epsilon_2)\,{\rm d}\epsilon_2,
\end{equation}
where $\mathscr R_{\rm pp}(x)$ is the angle-averaged pair production
rate ($\rm cm^3/s$) and reads 
\begin{equation}\label{eq:pairprodrate}
\begin{array}{lcl}
     \mathscr R_{\rm pp}(x)
     &=& c\displaystyle\int_{-1}^{\mu_{\rm crit}}{\rm d}\mu\,\frac{1-\mu}{2}\sigma(\beta),\quad \mu_{\rm crit}=\max(-1,1-2/x),\\
     &=& \displaystyle\frac 34\dfrac{c\sigma_{\rm Th}}{x^2}\displaystyle\,\psi\left(
     \dfrac{1+\sqrt{1-1/x}}{1-\sqrt{1-1/x}}
     \right)\Theta(x-1),
\end{array}
\end{equation}
and introducing the function
\begin{multline}\label{eq:psi}
   \psi(u) = -\dfrac 12 \ln^2(u) + \\
   +\left[
   \dfrac{2u(2+u)}{(u+1)^2}+\dfrac u4-\dfrac{2u}{u+1}+2\ln(1+u)-\dfrac 
12 +\dfrac 1{4u}
   \right]\ln(u)+ \\
   +2{\rm dilog}(u+1)-\dfrac u2 + \dfrac 1{2u} - \dfrac 2{1+u}+1+\frac {\pi^2}6
\end{multline}
and the dilogarithm function as \citep{as64,gould67a}
\begin{equation*}\label{eq:dilog}
\begin{array}{lcl}
   {\rm dilog}(x)
   &=&
   \DS-\int_1^x\D u\, \dfrac{\ln(u)}{(u-1)}\\
   &=&\DS -\frac {\pi^2}6 - \frac 12 \ln^2(x-1) + \sum_{k=1}^{\infty} \frac{(-1)^{k-1}}{k^2} (x-1)^{-k},\quad x>1.
\end{array}
\end{equation*}
To compute the photon escape probability $\mathscr P_{\rm esc}(\nu_1;t)$
of a photon of energy $\epsilon_1=h\nu_1/m_ec^2>1$ (or also the the
spectrum attenuation coefficient $\mathscr C_{\rm abs}(\nu_1;t)$), we
use the following approximate expression \citep{marcowith95},
\begin{equation}\label{eq:proba}
  \mathscr P_{\rm esc}(\nu_1;t)=\mathscr C_{\rm abs}(\nu_1;t)=\left[
\dfrac {1-\EXP{-\tau_{\gamma\gamma}}}{\tau_{\gamma\gamma}}
\right]
\,\EXP {-\tau_{\gamma\gamma}},
\end{equation}
with $\tau_{\gamma\gamma}\sim R\,{{\rm d}\tau_{\gamma\gamma}}/{\rm d\ell}$.
The factor between brackets is the usual solution of the transfer
equation in the plane-parallel geometry approximation, and can
approximate the photon escape probability in a spherical source of size
$R$. The extra exponential factor in equation (\ref{eq:proba})  has been
introduced by \citet{marcowith95} to account for the possibility for
high energy photons to annihilate outside the source, because the soft
target photons are not confined in the source like the particles; rather
their density decreases slowly on typical length scale equal to the
source radius.\\
Secondly, the high energy photons can also interact with the photon
field of the Diffuse Infrared Background (DIrB) (also called CIB for
Cosmic Infrared Background) during their travel through the universe
from the source to the observer \citep{gould67b}. Hereafter, we call
this effect \emph{extrinsic absorption}.  DIrB is the extra-galactic
light from the optical to sub-millimeter range, which records basically
the history of star formation \citep[for a review see][ and references
therein]{hauser01}. If we ignore the secondary gamma-ray emission in the
direction of the observer, it results that the emitted differential flux
is attenuated by a factor \citep{gould67b,stecker92,vassiliev00}
\begin{equation}\label{eq:probaext}
	\mathscr C_{\rm abs}^{\rm ext}(\nu,z_s)=
	\exp
	\left[-\tau_{\gamma\gamma}^{\rm ext}
	\left(
	\epsilon_1,z_s
	\right)
	\right],\qquad \epsilon_1=h\nu /  m_{\rm e}c^2.
\end{equation}
For close sources ($z_s<\!\!<1$), the expression (\ref{eq:taugg}) gives:
\begin{equation}\label{eq:tauggext}
	\tau_{\gamma\gamma}^{\rm ext}(\epsilon_1,z_s)\sim
	\dfrac{cz_s}{H_0}
	\int_{-1}^{1}\D\mu\ \dfrac{1-\mu}{2}\int_{\scriptstyle \epsilon_{\rm
	th}}^\infty \D \epsilon_2\, \sigma(\beta)
	n_{DIrB}(\epsilon_2),
\end{equation}
where $\epsilon_{\rm th}=2/(1-\mu)\epsilon_1$, $H_0$ is the Hubble
parameter (assumed equal to $65\ \rm km\,s^{-1}\,Mpc^{-1}$ throughout
this paper), and $n_{\rm DIrB}$ the density of the DIrB photon field. We
have estimated this density by performing a Chebyshev interpolation
using the measurements data compiled by \citet{hauser01} excluding the
two points of COBE-DIRBE at $60\ \mu\rm m$ and $100\ \mu\rm m$. But note
that with or without these points, the resulting absorption coefficients
are quite similar above 9 TeV as shown in figure \ref{fig:taugg} and do
not change the main results of this work.
\subsection{Pair production rate and pair cascade}\label{sec:pprat}
As mentioned above, the kinetic equation source term includes the 
contribution of the population of created particles in the pair 
production process as calculated above.
Remarking that a hard photon with a reduced energy $\epsilon>1$
interacts preferentially with a soft photon of energy  $\sim
1/\epsilon<1$ to form a pair $\rm e^+/e^-$ close to the production
threshold. Consecutively, both particles have thus a similar energy
$\gamma$ and we formally write the conservation of energy as
$\epsilon+1/\epsilon \approx \epsilon=2\gamma$. Then the pair production
rate reads
\begin{equation}
  Q_{\rm prod}(\gamma;t)=\frac{\D N}{\D t \D\gamma \D V}
  \approx 2 \dot{n}_{{\rm abs}} (
  \varepsilon = 2 \gamma ;t).
  \end{equation}
Assuming the IC emission is isotropic in the plasma rest frame,  the
differential photon absorption rate density per energy and time unit, is
given by
\begin{equation}
   \dot{n}_{{\rm abs}} ( \epsilon ) =
   4 \pi m_e c^2 \frac{j_{\nu}}{h^2 \nu}\mathscr P_{{\rm abs}} ( \nu;t ),
\end{equation}
and we finally obtain
\begin{equation}
   Q_{\rm prod}(\gamma;t)=
   \frac{8 \pi m_e
   c^2}{h^2}
   \left[
   \frac{j_{\nu}}{\nu}\mathscr P_{{\rm abs}} ( \nu;t )
   \right]_{\nu = 2 \gamma m_e c^2 / h}.
\end{equation}
\subsection{Time-averaged spectra}
At time $t$, the whole specific intensity in the plasma rest frame reads,
\begin{equation}
\mathscr I_\nu(\nu;t)\sim R \left[j_S(\nu;t)+j_C(\nu;t)\cdot \mathscr C_{\rm
abs}(\nu;t)\right]
\cdot \mathscr C_{\rm abs}^{\rm ext}(\nu,z_s),
\end{equation}
where all parameters and emission coefficients are expressed above. All
the physical quantities must be converted from the blob frame to the
observer frame, taking into account the Doppler boosting effect and the
cosmological corrections according
\begin{equation}
\tilde{\mathscr 
F}_{\tilde\nu}(\tilde\nu;t)\sim\pi\dfrac{R^2}{d_{\ell}^2}\delta_B^3(1+z_s)\mathscr 
I_{\nu}(\nu;t)
\text{ with }
\nu=\frac{1+z_s}{\delta_B}\,\tilde\nu,
\end{equation}
where $d_{\ell}$ is the usual luminosity distance,
$\delta_B=1/\Gamma_B(1-\beta_B\cos\theta)$ is the Doppler beaming factor
of the source and $\theta$ the viewing angle. \\
The observed spectrum is finally obtained by assuming that an
observation takes place in the interval $[\tilde t_{\rm obs},\tilde
t_{\rm obs}+{\Delta \tilde t}_{\rm obs}]$ (time $t=0$ is related to the
beginning of the injection of fresh particles). The time-averaged
spectrum is then
\begin{equation}
\tilde{\mathscr F}_{\tilde\nu}^{\rm obs}(\tilde\nu)=
\langle\tilde {\mathscr F}_{\tilde\nu}(\tilde\nu)\rangle_{\tilde t}
={1 \over \tilde{\Delta t}_{\rm
obs}}\int_{\tilde t_{\rm obs}}^{\tilde t_{\rm obs}+\tilde{\Delta
t}_{\rm obs}} \tilde{\mathscr F}_{\tilde\nu}(\tilde\nu;\tilde t) \ 
{\rm d}\tilde t.
\end{equation}
\section{Results and discussion}
\subsection{General behavior}
Our model requires eight parameters. Three of them are related to the
properties of the source, namely the magnetic field strength $B\ \rm$,
the radius $R$ (or $R_{15}$ when expressed in unit of $\rm 10^{15}\,
cm$) and the bulk Doppler factor $\delta_{B}$. Three others characterize
the injected plasma: they are the characteristic Lorentz factor of the
`pile-up' EDF $\gamma_{\rm max}$, the number of injected particles which
can be characterized by the integrated Thomson optical depth $\tau_{\rm
Th}=R\sigma_{\rm Th}\int\D t \int\D \gamma Q_{\rm inj}(\gamma;t)$ and
the injection time $t_{\rm inj}$. The remaining parameters are the
observational ones, ${t}_{\rm obs}$ and ${\Delta t}_{\rm obs}$.\\
In a steady-state model, there would only be five parameters since the
last three ones would be irrelevant. More exactly the integrated Thomson
optical depth $\tau_{\rm Th}$ should be replaced by the constant
particle optical depth $\tau_{\rm Th}^s = n R \sigma_{\rm Th } $, where
$n$ is the particle density in the source. For a `pile-up' distribution,
and neglecting pair creation,  the whole spectrum is entirely
characterized by two peak energies and two corresponding fluxes,
corresponding respectively to the synchrotron and the IC bumps.  Thus
there would be only one free parameter left, which can be taken for
instance as the unknown Doppler factor $\delta_B$ .  Exactly the same
spectrum would be obtained by varying $\delta_B$ and  adjusting the
other parameters accordingly. Since the TeV emission is  dominated by
the Klein-Nishina cut-off, the IC peak energy is simply proportional to
$\delta_B \gamma_{\rm max}$, whereas the synchrotron peak energy is
proportional to $\delta_B \gamma_{\rm max}^2 B$. Thus the following
scaling laws would apply :
\begin{equation}
\begin{array}{rcl}
	\gamma_{\rm max} 	&\propto& \delta_B^{-1},\\
	B 					&\propto& \delta_B.       \label{eq:bdelta}
\end{array}
\end{equation}
The integrated synchrotron luminosity scales as  $N \gamma_{\rm max}^2
B^2 \delta_B^4$, where $N = 4\pi R^3 n/3 = 4 \pi \tau_{\rm Th}^sR^2 /(3
\sigma_{\rm Th })$ is the total number of particles in the source. So
one gets the other following scaling law :
\begin{equation}
R^2 \tau_{\rm Th}^s \propto \delta_B^{-4}.
\end{equation}
The final condition must be determined by the fact that the IC
luminosity $L_{\rm IC}$ is directly related to the synchrotron photon
density in the source, which is itself directly related to the radius of
the source (the synchrotron luminosity does not depend on this radius
for a given number of particles).  The magnetic energy density scales as
$B^2$ so as $\delta_B^2$ (equation (\ref{eq:bdelta})), whereas the
synchrotron photon energy density scales as $L_{\rm IC}\delta_B^{-4}
R^{-2}$. If one neglects the Klein-Nishina correction, the ratio of
synchrotron to Compton luminosity is simply the ratio of magnetic to
soft photon density,and is fixed by the observations. One would thus
expect the following scaling
law
\begin{equation}
R \propto \delta_B^{-3}.
\end{equation} 
In fact the real condition is more involved because the Klein-Nishina
cut-off diminishes strongly the number of photons effectively available
for IC scattering, in a way depending on the Doppler factor and the
shape of the synchrotron spectrum. But qualitatively, one can always
choose the radius of the source to adjust the IC luminosity to the
observed value. \\
There is however a limitation to the Doppler factor due to the existence
of pair creation process. For to small values of $\delta_B$, the
$\gamma-\gamma$ optical depth, can increase so much that it becomes of
the order of unity. Then the IC luminosity stops increasing, and  rather
it starts to decrease with decreasing radius because of $\gamma-\gamma$
absorption. For a given Doppler factor, there is thus a maximum
reachable IC luminosity. Conversely for a given IC luminosity, there is
a minimum Doppler factor (which can be of course 1 in some cases where
pair creation is never important). Note that the variability time scale
is also a possible limitation, rapid variability requiring also high
Lorentz factors. \\
In principle the time-dependent model is more constrained. It requires
three more free parameters (the injection time and the two observation
times), but the entire shape of the synchrotron spectrum is depending on
these times. One can see that by varying $\delta_B$, the cooling time of
particles with the typical energy $ \gamma_{\rm max}$ varies as $(
\gamma_{\rm max} B^2)^{-1} \propto \delta_B^{-1}$ in the blob frame, so
as $\delta_B^{-2}$ in the observer frame. The shape of the synchrotron
spectrum will remain unchanged if one scales all times proportionally to
$\delta_B^{-1}$ in the blob frame, or $\delta_B^{-2}$ in the observer
frame. However, one of these times, namely the observation lasting time
${\Delta \tilde t}_{\rm obs}$ is not a free parameter. Thus if the model
fits perfectly well the data for all values of $\delta_B$, only one of
these values is compatible with the actual value of  ${\Delta t}_{\rm
obs}$. So theoretically a unique set of parameters (if any) can fit the
data.  Of course things are not so ideal :  because of observation error
bars, data will be fitted by a set of possible values with a
satisfactory $\chi^2$ test. 
\subsection{Approximated analytical solution of the kinetic equation}
\label{ap:analytical}
For illustrative purposes, we develop here the simplest case where 
one can neglect the Inverse Compton cooling in comparison
with the synchrotron cooling, and where pair production is 
unimportant.
We can express analytically
the general solution of equation (\ref{eq:kin}) which satisfies the boundary
condition $n_{\pm}(\gamma;t=0)=0$ for any arbitrary injection term
$Q_{\rm inj}(\gamma;t)$ by,
\begin{equation}\label{eq:formalsol}
n_{\pm}(\gamma;t)
=\dfrac{1}{\left|\dot\gamma(\gamma)\right|}\int_{\gamma}^{\infty} {\rm
d}\gamma_0\, Q_{\rm inj} \left(\gamma_0;t-\tau(\gamma_0,\gamma)\right),
\end{equation}
where $\tau(\gamma_0,\gamma)$ is the energy drift time, \emph{i.e.}
the time needed for a particle of energy $\gamma_0$ to cool down to
the energy $\gamma$,
\begin{equation}\label{eq:drifttime}
   \tau(\gamma_0,\gamma)= \int_{\gamma_0}^{\gamma} \dfrac{{\rm 
d}u}{\dot\gamma(u)}.
\end{equation}
For the injection term chosen as in equation (\ref{eq:injectionterm}), 
equation (\ref{eq:formalsol}) can be analytically integrated and we 
finally obtain,
\begin{equation}\label{eq:edf}
n_{\pm}(\gamma;t) =\dfrac {n_0\gamma_{\rm max}^3}{k_{\rm syn}\gamma^2}
\left[
\Gamma\left(3,{a(\gamma;t)\over\gamma_{\rm max}}\right)-
\Gamma\left(3,{b(\gamma;t)\over\gamma_{\rm max}}\right)
\right],
\end{equation}
where $\Gamma(\cdot,x)$ denotes to the incomplete gamma function.\\
Here the parameters $a(\gamma;t)$ and $b(\gamma;t)$ represent
respectively the lower and  upper bounds of the integral
(\ref{eq:formalsol}), where the integrand does not vanish. To evaluate
them,  we distinguish three time intervals for each value of $\gamma$. 
Let us define the
parameter $t_{\rm cool}(\gamma)=|{\rm d}\gamma/{\rm
d}\dot\gamma|=1/\gamma k_{\rm syn}$. Note that $t_{\rm
cool}(\gamma)=\tau(\infty,\gamma)$, {\itshape i.e.} it represents also
the time spent by an initial infinite energy particle to cool down to
$\gamma$.\medskip\\
{1.}\hskip 1em {\itshape Initial stage} where $t<\min(t_{\rm 
cool}(\gamma),t_{\rm inj})$
   \begin{equation*}
	b(\gamma;t)=\dfrac \gamma{1-k_{\rm syn}\gamma t}\quad;\quad
	a(\gamma;t)=\gamma.
   \end{equation*}
Particles are still injected at the energy $\gamma$ but high energy
particles have not yet cooled down to $\gamma$. Particles injected 
between $\gamma$ and some finite upper bound contribute to the 
integral. \\
{2.}\hskip 1em {\itshape Cooling stage} where $\max(t_{\rm 
inj},t_{\rm cool}(\gamma))<t<t_{\rm inj}+t_{\rm cool}(\gamma)$
\begin{equation*}
     b(\gamma;t)\to \infty\quad;\quad
     a(\gamma;t)=\dfrac \gamma{1-k_{\rm syn}\gamma (t-t_{\rm inj})}.
\end{equation*}
Particles are no more injected but some high energy particles are
still cooling down to $\gamma$. Particles injected above a finite 
energy (larger than $\gamma$) contribute to the integral.\\
{3.}\hskip 1em {\itshape Intermediate stage.} For intermediate values of
$t$, we must distinguish two specific energy ranges. We define a
critical value of the individual Lorentz factor of the particles,
$\gamma_{\rm lim}=1/k_{\rm syn}t_{\rm inj}$:\smallskip\\
\null\hskip 1em 3.a)\hskip 1em {\itshape low energy range} where
$\gamma<\gamma_{\rm lim}$ (or $t_{\rm inj}<t_{\rm cool}(\gamma)$),
\begin{equation*}
     b(\gamma;t)=\dfrac \gamma{1-k_{\rm syn}\gamma t}\quad;\quad
     a(\gamma;t)=\dfrac \gamma{1-k_{\rm syn}\gamma (t-t_{\rm inj})}.
\end{equation*}
Injection is finished but very high energy particles have not yet
cooled down to $\gamma$. Particles injected in some interval above 
$\gamma$ 
 contribute to the integral. \smallskip\\
\null\hskip 1em 3.b)\hskip 1em {\itshape high energy range} where
$\gamma\geqslant\gamma_{\rm lim}$ (or $t_{\rm inj}\geqslant t_{\rm 
cool}(\gamma)$),
\medskip
\begin{equation*}
     b(\gamma;t)\to \infty\quad;\quad
     a(\gamma;t)=\gamma.
\end{equation*}
Conversely, very high energy particles have time to cool down to
$\gamma$ while injection of fresh particles still takes place. All
particles injected above $\gamma$ contribute. Note that this is the only
stage for which  $n_{\rm}(\gamma;t)$ does not depend on time.  A
steady-state is set during this stage (although not in the spectrum
because the whole distribution is not steady).

We could also define an {\itshape end stage} for which $t>t_{\rm
cool}(\gamma)+t_{\rm inj}$ and where $n_{\pm}(\gamma)=0$.

\begin{figure}
\plotone{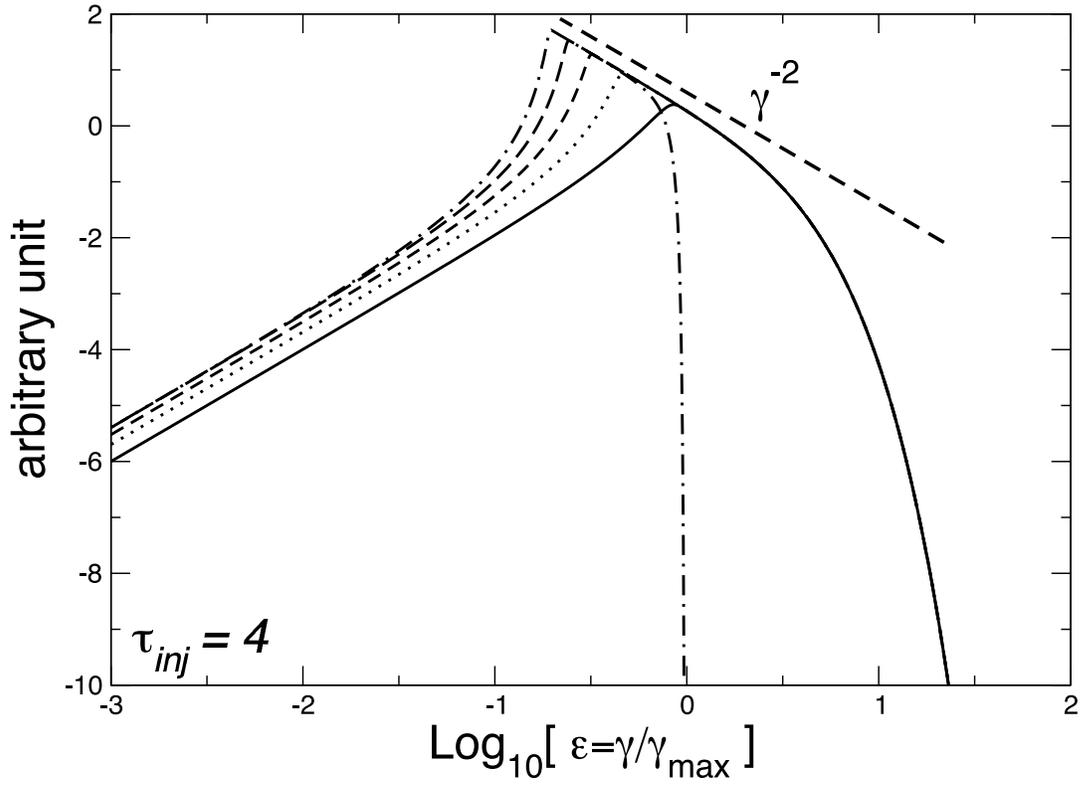}
\caption{%
         An example of resulting cooling `pile-up' energy distribution
         function $\varpi_{\tau_{\rm inj}}(\varepsilon;\tau)$ for
         $\tau_{\rm inj}=4$ a time   $\tau=1,2,3,4,5$. Also represented,
         in dashed straight line, a power-law of index 2 typical which
         results from the radiative cooling of a mono-energetic particle
         energy distribution.
}\label{fig:EDF}
\end{figure}

Introducing the reduced variables, $\varepsilon={\gamma}/{\gamma_{\rm 
max}}$ and $\tau_i={t_i}/{t_{\rm cool}(\gamma_{\rm max})}$, the 
previous equations can be collected into the following relation,
\begin{equation}
n(\varepsilon;\tau)=\dfrac {n_0\gamma_{\rm max}^3}{k_{\rm syn}\gamma^2}
\,\varpi_{\tau_{\rm inj}}(\varepsilon;\tau),
\end{equation}
where
\begin{multline}
\varpi_{\tau_{\rm 
inj}}(\varepsilon;\tau)=\varepsilon^{-2}\Theta(1-\varepsilon 
\max(0,\tau-\tau_{\rm inj}))\times\\
 \quad \times \left[
\Gamma\left(3,\dfrac{\varepsilon}{1-\varepsilon \max(0,\tau-\tau_{\rm 
inj})}\right)-\Theta(1-\varepsilon\tau)\Gamma\left(3,\dfrac{\varepsilon}{1-\varepsilon\tau}\right)
\right].
\end{multline}
An example of a resulting cooling pair-EDFs at different times is
plotted in figure \ref{fig:EDF}. One clearly sees the initial stage
where the EDF is built, the formation of a $\gamma^{-2}$ EDF due to the
cooling and the subsequent cooling of the whole distribution after the
injection has stopped.  As we will see in realistic simulations, the
shape is however strongly modified when taking into account the IC
cooling process (which is not simply dependent on the energy because of
the Klein-Nishina cut-off) and the pair production term.
\subsection{Application to Mrk 501  data}
\subsubsection{Observations}
We have applied the model to fit the spectral energy distribution of
Mrk\ 501 during the period of 1997 April when this source experienced an
intense period of activity. From this period, we distinguish two
different activity states, namely the `high state' from the April 16 and
the `medium state' form the April 7.  Simultaneous data are taken from
BeppoSAX for the X-ray observation \citep{pian98} and from the French
Atmospheric {\vC}erenkov Telescope {\vC}AT for the spectrum in the TeV
energy regime \citep{cat99,barrau98}.
In a first step, we have corrected the high energy spectra using the
attenuation coefficient computed previously. Note that the last
corrected data point of the high state may be not meaningful, leading to
a concave up corrected high energy spectrum. The most simply and obvious
explanation of this problem is an over estimation of the measured high
energy tail of the blazar spectrum or/and of the DIrB density itself.\\
The value of the $\Delta \tilde t_{\rm obs}$ is constrained by the
observing duration. However there are some discontinuities in the
observing time, which make the global observation period length
different from  the real exposure time. We have assumed that on average
our $\Delta \tilde t_{\rm obs}$ corresponds to the global observing time
of the BeppoSAX instruments LECS/MECS and PDS ($\sim$40000 s for the
April 16, 37845 s for the April 7). \\
\subsubsection{Solutions without pair creation}
\begin{figure}
\plotone{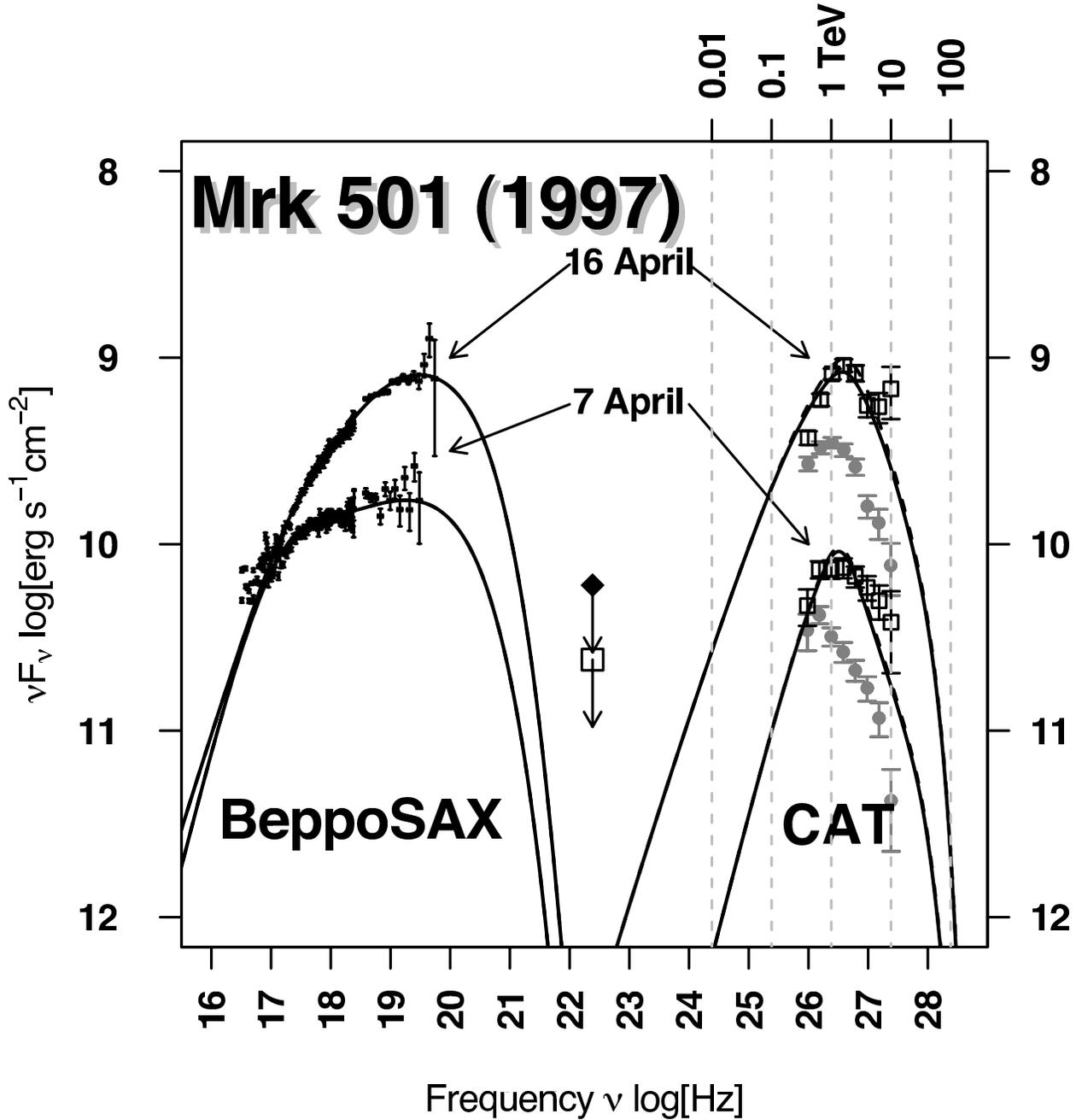}
\caption{%
         Fits of Mrk 501 data during the flaring period in 1997 April by
         the model.  See Table \ref{tab:parameters} for the detail of
         the parameters values.  For each state, the solid curves show
         the emergent spectrum modified by the intrinsic absorption due
         to the pair creation process in the spherical blob. The dashed
         curves show the unabsorbed one. Filled gray circles : observed
         {\vC}AT data. Open squares : data corrected from the DIrB
         attenuation.
}\label {fig:fit}
\end{figure}

We can reproduce the data with parameters corresponding to almost no
pair creation.  The parameters used to fit the data are reported in
Table \ref{tab:parameters} and  the resulting synthetic SED are plotted
in Figure \ref{fig:fit}. The fits are quite satisfactory for the high
energy part of the SED. One-zone models are appropriate only to to
reproduce this high energy, variable part.  They can not account for the
radio emission produced at much larger scale,  where the whole jet
contributes, possibly being the superposition of  many successive
flaring events.\\
In some range of Lorentz factor, a steady-state solution corresponding
to $\gamma^{-2}$ power-law (see equation (\ref{eq:edf}), case 3.b)) can
be observed giving a $\nu^{-1/2}$ synchrotron flux index. This
corresponds to the main part of the April 16 spectrum and and the low
energy part of the April 7 one. Above this range, the spectrum is
modified by a factor $t_{\rm cool}/t_{\rm obs}$ because the particles
have cooled before the end of the observation. This produces a
steepening of the spectrum by $\Delta\alpha=0.5$ and explains the flat
high energy part of the synchrotron spectrum. The position of the
spectral break is thus directly determined by the ratio between the
cooling and the observation time.
We obtain values of the magnetic field and of the transverse radius of
the source ($R_{15}\sim 1$) in agreement with other models
\citep[see \textit{e.g.}][
]{bednarek99,tavecchio01,ghisellini02,katarzynski01,konopelko03}. It
turns out that this minimum Doppler factor implied by $\gamma-\gamma$
absorption in steady state models is quite high for Mrk 501, as noticed
already by several authors. When the IC luminosity is corrected from
extrinsic absorption, it can be as high as 50 for stationary models  
\citep[see \textit{e.g.}][]{bednarek99,konopelko03}.
In the time dependent model, the
constraint is somewhat released because high energy photons (TeV) are
not produced at the same time as low energy, IR photons that are mainly
responsible for their absorption. So the actual density of soft photons
during the emission of highest energy TeV photons is lower than that
measured on average. We see that good fits can be obtained with Doppler
factors around 25, which implies a bulk Lorentz factor larger than 12. 
Not that this still relatively high value is necessary due to the
correction of the spectra from from the DIrB absorption.  These high
values are generally in conflict with other features of TeV BL Lacs ;
first VLBI/VLBA sub-parsec observations of Mrk 501 and others TeV
blazars which don't exhibit superluminal velocities at mas scale or
excess in the derived synchrotron brightness temperature
\citep{edwards02,piner04}. Moreover, if we suppose that BL Lacs are the
beamed counterparts of the FR-I radio galaxies, the ratio of their
bolometric luminosity should be of the order of $\delta_B^8$. A
$\delta_B$ larger than 10 would result in a much larger contrast than
what is observed \citep{urry95}. The spatial density of beamed vs
unbeamed objects seems also to be better consistent with a Lorentz
factor of around 3. All these argument disfavour a high Lorentz factor.
This is in fact a problem inherent to any one zone model. We will argue
in a future work (Saug\'e \& Henri in preparation) than inhomogeneous 
models may solve this issue.

Like $B$ and $\delta_B$, $\gamma_{\max}$ is kept constant between the
two different states.  In the light of our scenario, the `medium state'
spectrum could be due to a previous ejection observed in a later stage
(with respect of the injection time) than the April 16 one and in a much
larger part of the jet.

We can also estimate the minimum variability time scale deriving from
standard causality arguments and given by $\tilde t_{\rm var}>\tilde
t_{\rm var,min}=555\,R_{15}\,(1+z_s)\delta_B^{-1}\ \rm min$. Its value
is reported in the last column of the Table \ref{tab:parameters} and is
equal to 15 min for the high state and roughly 40 min for the medium
state where the size of the source is much larger as mentioned above.
Note however that the injection time is much larger than the above
values.  

\begin{figure}
\plottwo{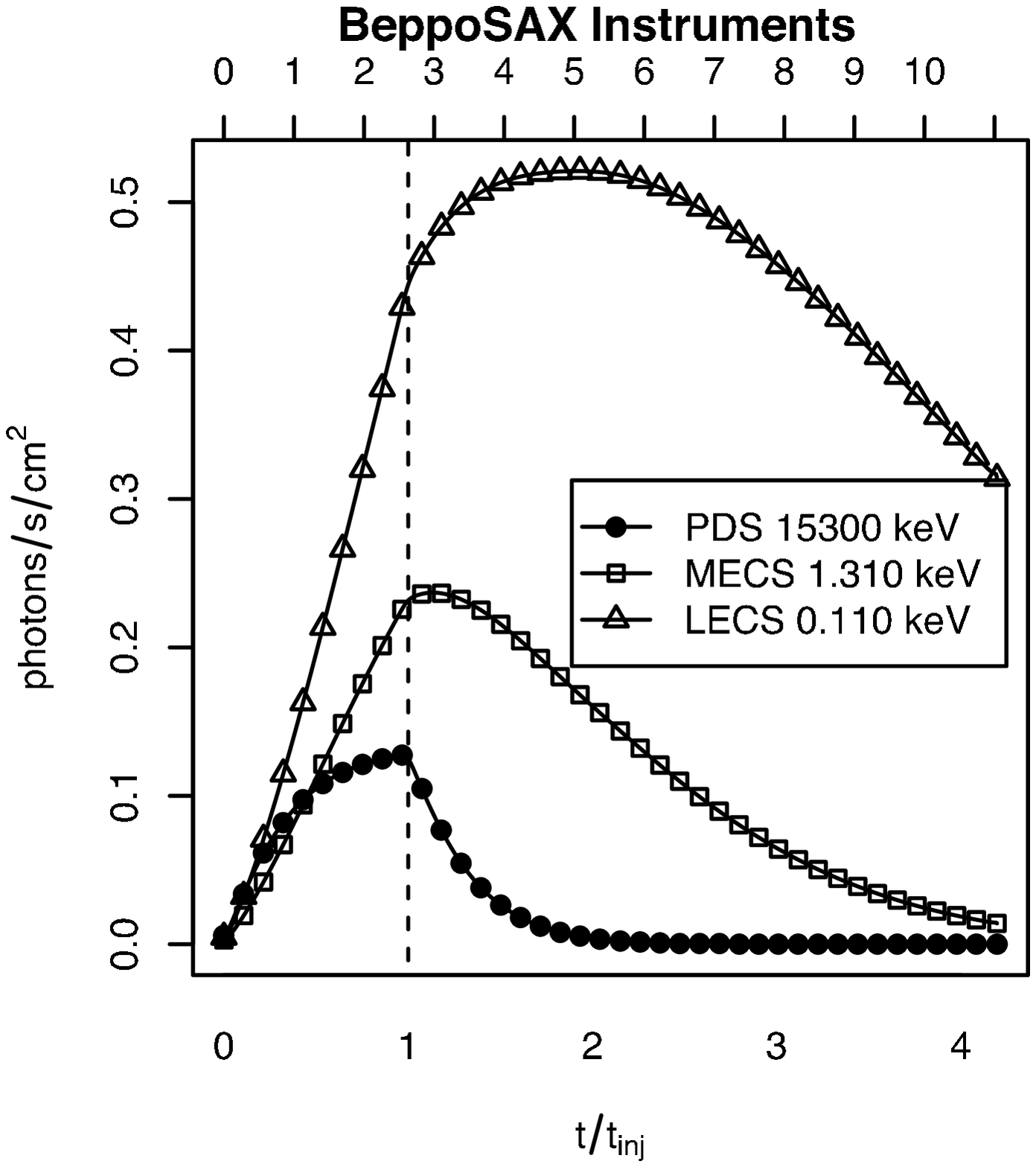}{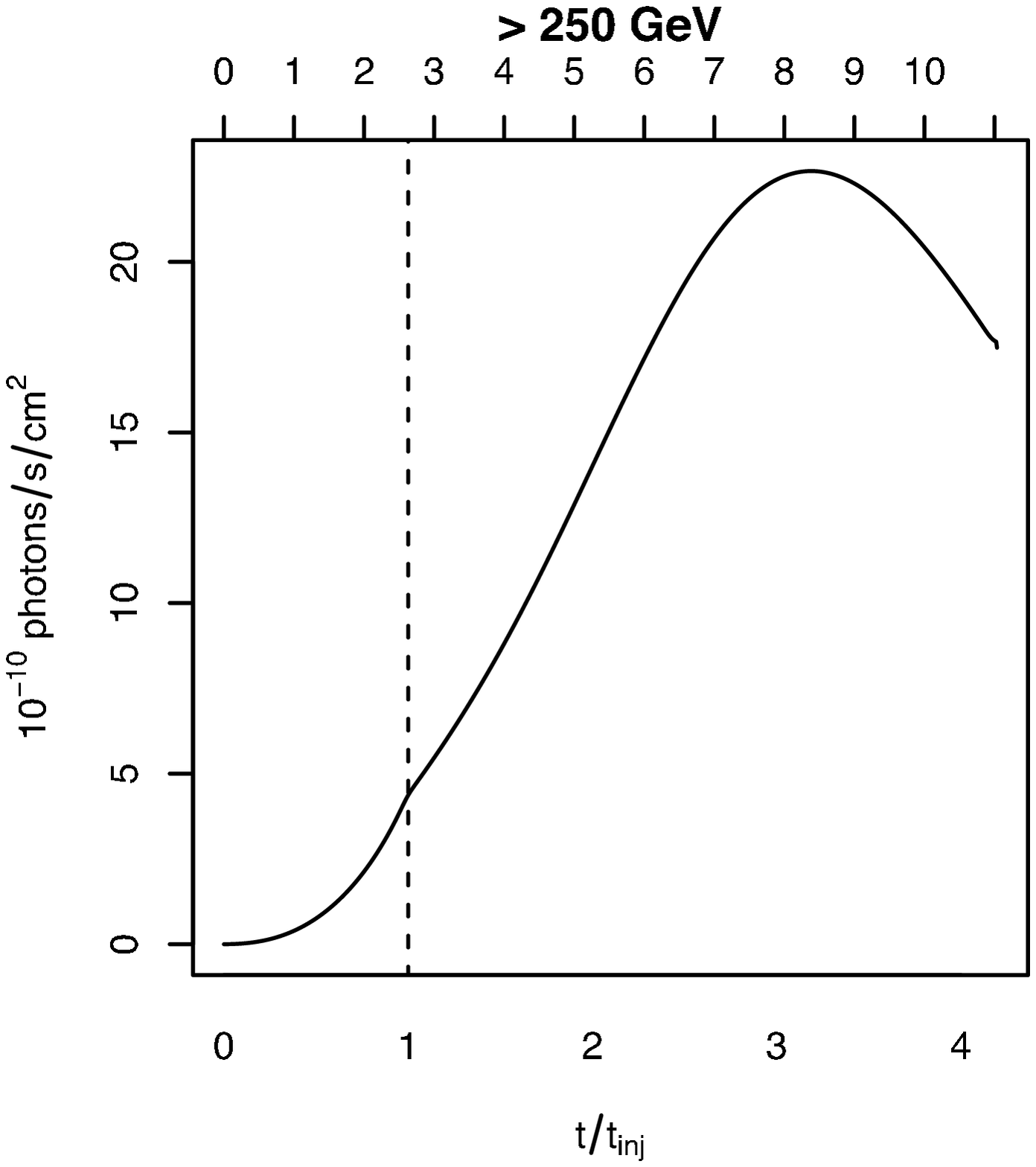}
\caption{%
         Mrk\ 501 April 16 simulated light curves for BeppoSAX
         instruments (left panel) and above 250 GeV (right pannel).
         Upper axis is graduated in unit of hours in the observer frame.
      }\label{fig:lightcurves1}
\end{figure}

Simulated light curves corresponding to April 16 parameters are shown in
figure \ref{fig:lightcurves1}. The curves are calculated for the 3
energy ranges of BeppoSAX and for very high energy (VHE) instruments
above 250 GeV. Lags between various energy bands are clearly visible on
the curves. Remarkably, the high energy photons lead the soft energy
ones in the X-ray synchrotron component. but the high energy gamma-ray
curve lags the synchrotron component. This is due to the fact that when
particle distribution cools, the photon density below the Klein-Nishina
limit increases. This makes the IC emission level keep rising even after
the injection has stopped. Precise comparisons with observed light
curves have not been made because the statistics is too poor to allow a
meaningful analysis. However we can see qualitatively that complex
temporal effects can arise from a time-dependent simulation leading to
apparently contradictory behaviours. Note that whereas the presented
simulations span the whole temporal interval of BeppoSAX observations,
high energy observations have been performed only during a limited time
within this interval. If they have taken place around the maximum of the
light curve, few variations are expected. 

Of course, a more realistic model could include a succession of
different flares as well as a variation of the injection characteristic
energy. This is however beyond the scope of this paper which only aims
at showing that `pile-up' particle energy distribution can well explain
the high energy shape of TeV blazar spectra.  
\subsubsection{Solutions with pair creation}\label{sec:gevflare}
\begin{figure}
\plotone{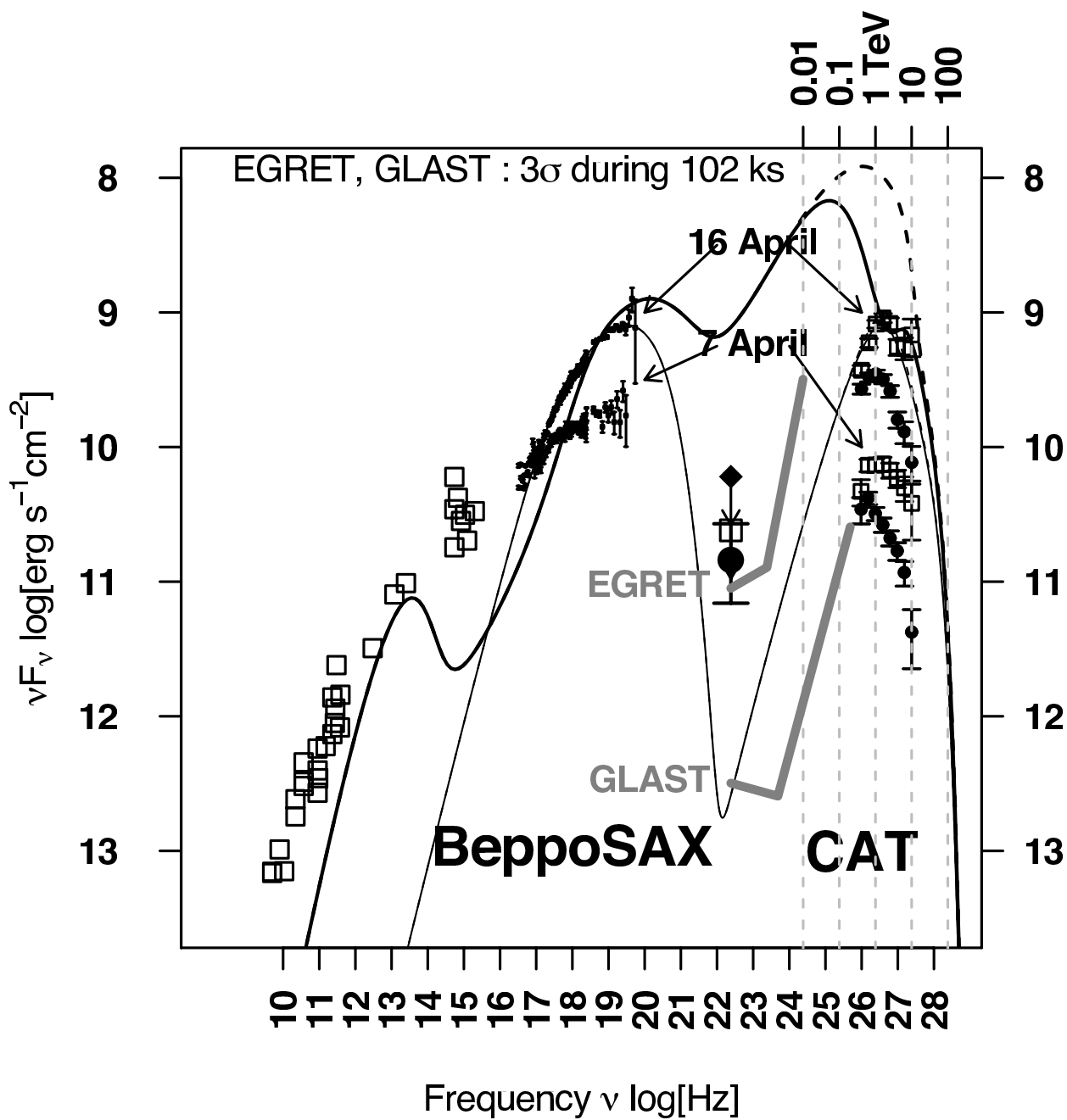}
\caption{%
         Time-averaged homogeneous one-zone SSC model for spectra of Mrk
         501 in the case of an intense pair production. Thick grey lines
         show the EGRET and GLAST instruments sensitivity curve
         (3$\sigma$ during 102 ks). 
}\label {fig:flash}
\end{figure}

\begin{figure}
\plottwo{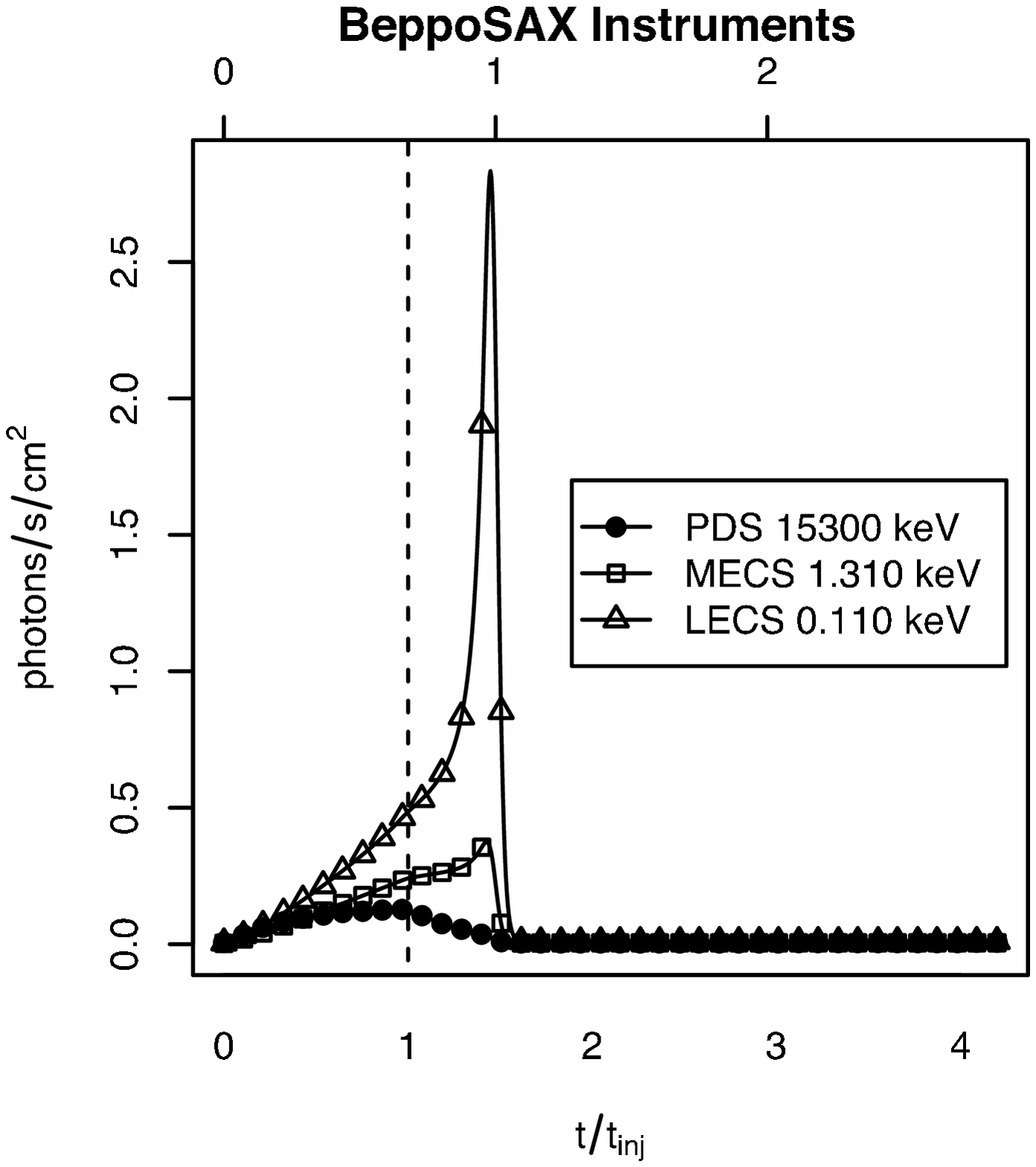}{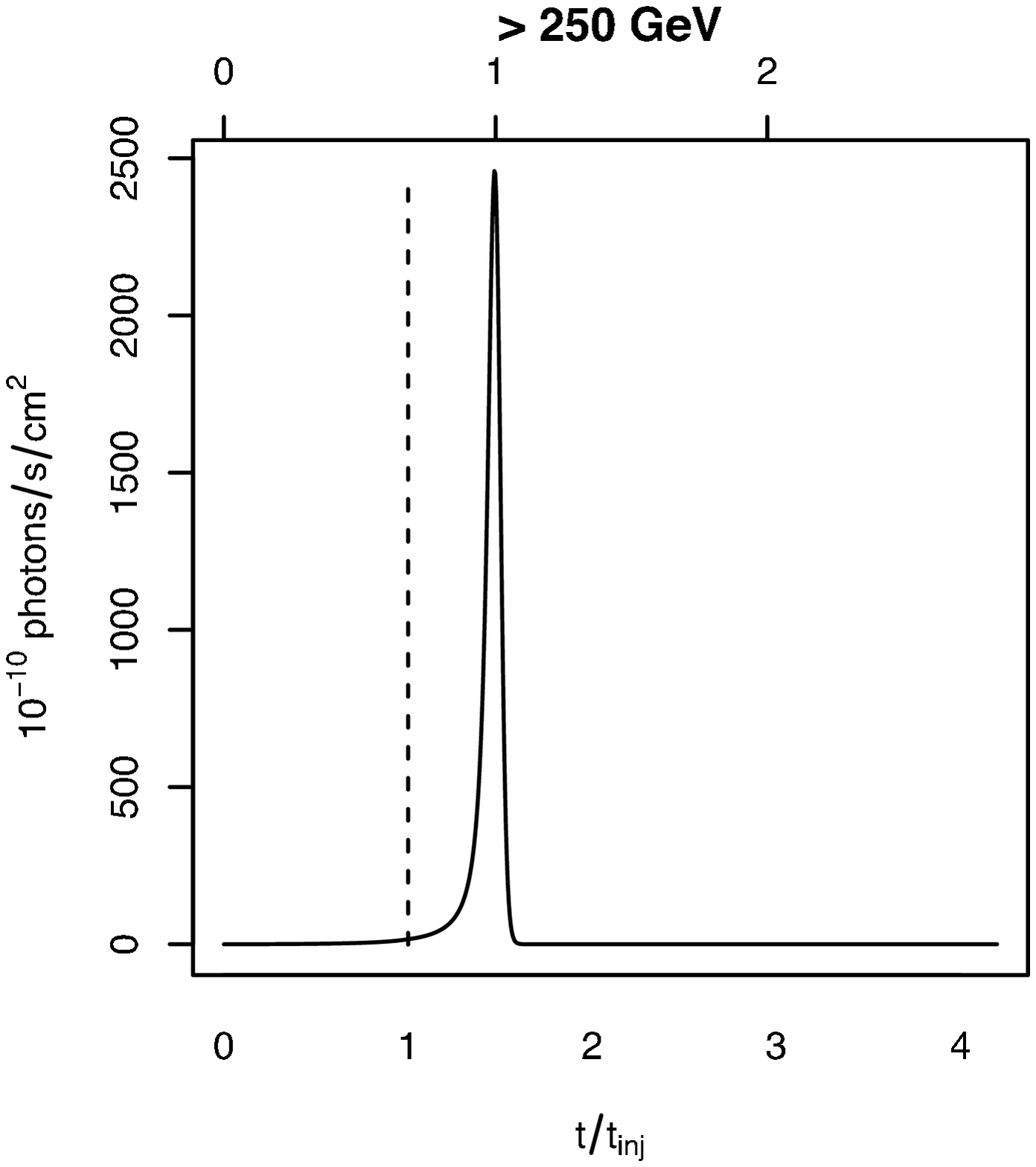}
\caption{%
         Same as figure \ref{fig:lightcurves1} for the case of an
         intense pair production.
         }\label{fig:lightcurves2} \end{figure}

Interestingly, we found also possible solutions with a much lower bulk
Doppler factor, and accordingly a lower magnetic field, a higher
particle Lorentz factor and a larger radius. An example of such
parameters is shown in Table 2, and corresponding spectrum is displayed
in figure \ref{fig:flash}. In this regime, pair production can be
important. Although the IC flux should be much larger, the
$\gamma-\gamma$  absorption reduces effectively the luminosity so that
it can be compatible with the observations. The intense creation of new
particles produces more synchrotron photons, which accelerates the
cooling. This phenomenon also amplifies the effect of the increase of
the Klein-Nishina threshold energy. This leads eventually to a
catastrophic pair production/cooling process, producing a strong flare
at GeV energy and in low energy X-rays. It explains the bump occuring in
the GeV/sub-TeV energy range (and to a lesser extent in the radio
sub-millimeter range). The flare is also more clearly visible on the
light curves (figure \ref{fig:lightcurves2}) appearing as a very sharp
flare in some energy ranges. Although the relevance of such scenario is
not  clear, it may be possible that such events could explain the most
rapid variations in the TeV light curves. The GeV flare would have been
in principle observable by the EGRET instrument. However due to the
small number of simultaneous EGRET/TeV observations and the briefness of
the event, it could have escaped any detection. We note also that for
different value of the parameters, it is possible that the X-ray flare
shifts toward lower energy, disappearing from the X-ray data. This could
be related to the `orphan flare events' observed in some occasions
\citep{kraw04}. 
\section{Conclusion}
This paper shows that the high energy spectra of TeV blazars can be well
reproduced by a cooling `pile-up' EDF. This offers an alternative to the
narrow power-law injection terms often used in the literature, whose
justification is unclear. Inclusion of time-dependent effects permits to
reproduced the main features of both the light curves and the
time-averaged spectra. For a given SED shape, the parameters of the
model are fully constrained. For Mrk\ 501, different states could be the
result of the time variation of the transverse radius of the source $R$,
the quantity of injected leptons via $\tau_{\rm Th}$ and the
observational parameters with respect of the injection time.  However,
this one zone model shares a common issue with other homogeneous models
: it requires high values of the bulk Lorentz factor to avoid a strong
gamma-ray absorption, even in the case of the optically thick solution
(see section \ref{sec:gevflare}).  These high Lorentz factors appear to
be inconsistent with those deduced from FR I/BL Lacs unification models
\citep{urry95,chiaberge00}. They are also difficult to reconcile with
the absence of superluminal motion at TeV scale and relatively low
brightness temprature of TeV blazars \citep{edwards02,piner04}.  We will
argue in a future work that inhomogeneous models could fix this issue. 
\acknowledgments We thank
A. Djannati-Ata{\"\i}, 
P. Espigat,
F. Piron, 
S. Pita and 
A. Lemi\`ere 
from the {\vC}AT collaboration and A. Marcowith for helpful interaction.
Critical remarks of an anonymous referee helped to improve significantly
the final version of this paper.  L.S.  would like to thank J. Ferreira
and P.-O. Petrucci for very enthusiastic and stimulating discussions and
encouragement. Part of the simulations reported here has been performed
at the ``\emph{Centre de Calcul Intensif de l'Observatoire de
Grenoble.}''.



\appendix
\section{An accurate analytic approximation of pitch angle averaged\\
synchrotron emitted power
}\label{ap:synch}
As mentioned in section \ref{sec:synchprocess}, the $R_{{\rm CS}}(x)$ function (see equation\ref{eq:RCS}) results from
the monochromatic emitted power averaged on a population of particles with
randomly distributed pitch angle. It mathematically reads \citep{crusius86}
\begin{equation}
   R_{{\rm CS}} ( x ) = \frac{1}{2} \int_0^{\pi} \D \theta\, \sin^2 \theta
   F_{{\rm syn}} \left( \frac{x}{\sin \theta} \right),
\end{equation}
where $F_{{\rm syn}} ( x )$ is the usual synchrotron fundamental kernel \citep{bg70},
\begin{equation}
   F_{{\rm syn}} ( x ) = x \int_x^{\infty} \D z\, K_{5 / 3} ( z ).
\end{equation}
For $x \ll 1$ approximate expression of $F_{{\rm syn}}$ is
\begin{equation}
   F_{{\rm syn}} ( x ) \approx \text{ $\frac{4 \pi}{\sqrt{3} \Gamma ( 1 / 3 )}
   \left( \frac{x}{2} \right)^{1 / 3}$} \quad,\quad x \ll 1.
\end{equation}
and we immediately obtain the relevant one for $R_{{\rm CS}}$,
\begin{equation}
   R_{\ll} ( x ) = \frac{1}{2} \int_0^{\pi} \D \theta\, \sin^2 \theta
   F_{{\rm syn}} \left( \frac{x}{\sin \theta} \right) = \frac{2^{1 / 3}}{5}
   \Gamma^2 ( 1 / 3 ) x^{1 / 3} \approx 1.808418\, x^{1 / 3} .
\end{equation}
Conversely, for large argument ($x \gg 1$), the asymptotic development of
$F_{{\rm syn}}$ is $\left. ( \pi x/2 \right)^{1/2} \mathe^{- x}$ and then
\begin{equation}
   R_{\gg} ( x ) = \frac{1}{2} \int_0^{\pi} \D \theta\, \sin^2 \theta
   F_{{\rm syn}} \left( \frac{x}{\sin \theta} \right) = \frac{1}{2} \left(
   \frac{\pi x}{2} \right)^{1 / 2} \int_0^{\pi} \D \theta\, \sin^{3 / 2}
   \theta \mathe^{- x / \sin \theta}.
\end{equation}
The preceding integral can re-writes as,
\begin{equation}\label{eq:integralI}
   \mathcal I ( x ) = \int_0^{\pi} \D \theta\, \exp ( - f_x ( \theta ) ) \quad, \quad
   f_x ( \theta ) = \frac{x}{\sin \theta} - \frac{3}{2} \ln(\sin \theta).
\end{equation}
Remarking that the exponential argument is symmetric around $\theta = \pi / 2$
and therefore on the integration range, we set $\theta
= \pi / 2 + \varphi$ and make a taylor expansion around $\varphi$
\begin{equation}
   \tilde{f}_x ( \varphi ) = \frac{x}{\cos \varphi} - \frac{3}{2} \ln(\cos
   \varphi) \approx x + \frac{\varphi^2}{2} \left( x + \frac{3}{2} 
\right) + \mathcal{O} (
   \varphi^3)
\end{equation}
Then integral (\ref{eq:integralI}) rewrites, 
\begin{equation}
   \mathcal I ( x ) \approx \mathe^{- x} \int_{- \pi / 2}^{\pi / 2} \D \varphi\,
   \mathe^{- \varphi^2 / 2 \sigma^2} \quad {\rm with} \quad 1 / \sigma^2 = x +
   \frac{3}{2},
\end{equation}
which is a standard gaussian integral. Noting that larger is $x$ sharper
is the integral argument, lower and upper bound can be extended to
infinity as $x\gg1$ (error is less than $10^{- 10}$ for $x \geqslant 3$
and ). We finally obtain,
\begin{equation}
   R_{\gg} ( x ) = \frac{\pi}{2} \sqrt{\frac{2 x}{x + 3}} \mathe^{- x}.
\end{equation}
For intermediate values, we perform a polynomial fit of the form,
\begin{equation}
   R_{{\rm int}} ( x ) = a_0 ( x^{a_1} + a_2 x^{2 a_1} + a_3 x^{3 a_1} ) \exp
   ( - a_4 x^{a_5} ),
\end{equation}
where the coefficient $a_i$ is obtained from least square fitting and is given
in table \ref{tab:fitcoeffs}. In this regime, RMS error is in order of 0.05\% .

\clearpage
\begin{deluxetable}{lccccccccc}
\tabletypesize{\scriptsize}
\tablecaption{Mkn 501 physical fit parameters\label{tab:parameters}}
\tablewidth{0pt}
\tablecolumns{9}
\tablehead{
   \colhead{} &
   \colhead{$B$} &
   \colhead{$R_{15}$} &
   \colhead{$\delta_{B}$} &
   \colhead{$\gamma_{\rm max}$} &
   \colhead{$\tau_{\rm Th}$} &
   \colhead{$\tilde t_{\rm inj}$} &
   \colhead{$\tilde t_{\rm obs}$} &
   \colhead{${\Delta \tilde t}_{\rm obs}$} &
   \colhead{$\tilde t_{\rm var,min}$} 
   \\
   \colhead{}&
   \colhead{(G)}&
   \colhead{$\rm\ (10^{15}\,cm)$}&
   \colhead{}&
   \colhead{$10^6$}&
   \colhead{$10^{-10}$}&
   \colhead{(ks)}&
   \colhead{(ks)}&
   \colhead{(ks)}&
   \colhead{(min)}
}
\startdata
high state
	& 0.077
	& 0.65
	& 25
	& 1.26
	& 58.5
	& 9.4
	& 3.29 ($= 0.35\, \tilde t_{\rm inj}$)
	& 39.7
      & 15
  \\
medium state
	& 0.075
	& 1.75
	& 25
	& 1.29
	& 8.95
	& 40
	& 35 ($= 0.875\, \tilde t_{\rm inj}$)
	& 37.8
    & 40
  \\
  \hline
`GeV flaring state' 
	& 0.047
	& 1.06
	& 16
	& 2
	& 144.1
	& 24.4
	& 8.54 ($= 0.35\, \tilde t_{\rm inj}$)
	& 102
    & 38
\enddata
\end{deluxetable}
\clearpage
\begin{table}
\centering
\caption{%
Resulting fitting coefficients from the polynomial approximation
of the function $R_{\rm CS}$\label{tab:fitcoeffs}}
\vskip 1em
\begin{tabular}{r@{ = }r@{.}l}
  	\hline\hline
       $a_0$ &  0&201 447$\times 10$\\
     	$a_1$ &  0&344 606\\
     	$a_2$ & -0&429 682\\
     	$a_3$ &  0&273 331$\times10^{-2}$\\
     	$a_4$ &  0&966 844\\
     	$a_5$ &  0&964 518 \\ \hline
       $\chi^2$   & 1&27$\times 10^{-5}$\\
     RMS \% error & 0&05\\
     \hline
\end{tabular}
\end{table}
\clearpage
\clearpage
\clearpage
\clearpage
\clearpage
\clearpage
\end{document}